# A VLSI Design Flow for Secure Side-Channel Attack Resistant ICs

Kris Tiri[1] and Ingrid Verbauwhede[1,2]

[1]UC Los Angeles, [2]K.U.Leuven
{tiri, ingrid}@ee.ucla.edu

## Abstract

*This paper presents a digital VLSI design flow to create secure, side-channel attack (SCA) resistant integrated circuits. The design flow starts from a normal design in a hardware description language such as VHDL or Verilog and provides a direct path to a SCA resistant layout. Instead of a full custom layout or an iterative design process with extensive simulations, a few key modifications are incorporated in a regular synchronous CMOS standard cell design flow. We discuss the basis for side-channel attack resistance and adjust the library databases and constraints files of the synthesis and place & route procedures accordingly. Experimental results show that a DPA attack on a regular single ended CMOS standard cell implementation of a module of the DES algorithm discloses the secret key after 200 measurements. The same attack on a secure version still does not disclose the secret key after more than 2000 measurements.*

## 1 Introduction

The integrated circuit (IC) emerges more and more as the main vulnerability of a security application. Due to physical and electrical effects, it broadcasts information that is related to the secret key. In recent years, several attacks have been reported that use information from so-called 'side-channels' to find the secret key. These side-channel attacks do not tamper the security IC. They are non-invasive and observe the device under normal operation mode. The attacks analyze information ranging from time delay [1] and power consumption [2] to electromagnetic radiation [3] and often apply statistical techniques. In general, SCAs do not require expensive equipment and are rather quick to set up. SCAs are a real threat for any device of which the security IC is easily observable such as smart cards and embedded devices.

At first, SCAs have been fought with ad hoc countermeasures. For instance, the addition of random power consuming operations obscured the data dependent variations in the power consumption. Yet over time, the attacks have evolved and become more and more effective. Subsequently, countermeasures have been conceived at the different abstraction levels of the security application. It started at the algorithmic level. A fine illustration is masking [4]. This technique prevents intermediate variables to depend on an easily accessible subset of the secret key. Only recently, dedicated hardware techniques have been presented [5],[6]. Instead of concealing or decorrelating the side-channel information, these techniques try not to *create* any side channel information. Goal of these countermeasures is to make the power consumption of the logic gates independent of the data values or the Hamming distances between data values. To our knowledge, this publication is the first to present a comprehensive top-down automated synchronous VLSI design flow that pursues a constant power dissipation of the security IC.

A major advantage of our proposed technique is that it is independent of the cryptographic algorithm or arithmetic implemented and that there is no need to train the VLSI designer to become a security expert or vice versa. When the power dissipation of the smallest building block is a constant and independent of the signal activity, no information is leaked through the power supply regardless of the implementation and consequently irrespective of the experience of the digital designer. Such a technique is an essential component of a secure digital design flow.

A secure digital design flow is an automated design flow that creates a secure integrated circuit or system-on-chip. The design flow starts from the design specifications and results in a secure side-channel attack resistant layout through the subsequent steps of synthesis and place & route. Major smart card vendors and service providers have recently identified such a design flow as an important open issue related to the general security of cryptographic applications [7].

In this paper, we will transform a regular synchronous digital design flow into a secure digital design flow. The modifications and additions are inserted in the backend of the regular automated design flow and have been implemented in a "push-button" approach. They only have a minimal influence on the design flow and a negligible overhead in design time. The additional steps required only a total of 6 minutes of extra CPU time for a 39K gates prototype IC implementing a high-throughput AES, controller and fingerprint processor.

The remainder of this paper is organized as follows. In section 2, we present the secure digital design flow. We

This work was supported in part by NSF grant CCR-0098361.

first discuss two components: (1) a logic style with constant power consumption and (2) a place & route technique, which controls the parasitic effects on the interconnect wires. Next, the remaining components, which complete the design flow, are presented. Section 3 compares the secure digital design flow with a regular digital design flow. A module of the DES algorithm is implemented. Area and power numbers are given and the results of a Differential Power Analysis are provided. While section 2 and 3 focus on power attacks, section 4 will deal with the other side-channel attacks. Finally a conclusion will be formulated.

## 2 Secure Digital Design Flow

The secure digital design flow is depicted in Fig. 1. Someone familiar with a regular synchronous CMOS standard cell based digital design flow recognizes the subsequent steps in an IC design: logic design, logic synthesis and place & route. The creative part is the description of the circuit in a high level design language such as Verilog or VHDL, indicated by the "logic design" task. Nothing changes to this task. One can notice 2 additional steps: "cell substitution" and "interconnect decomposition". These operations have been inserted in the backend of the flow and do not interfere with the creative part of a design. Before we discuss each step, we first elaborate on 2 important components, which are necessary to understand the flow. We will start with a logic style with constant power consumption. The idea is to have exactly 1 switching event per clock cycle during which a constant amount of charge is used. A 100% switching factor can only be achieved through a dual rail logic, which has a true and a false output. Consequently, the load capacitance at the 2 outputs must be matched to assure that the load capacitance is independent of the switching event. This is achieved through a special place & route approach, which we will discuss secondly.

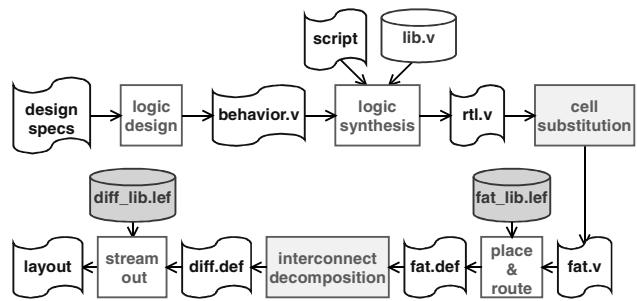

**Fig. 1. Main tasks of secure digital design flow.**

### 2.1 Wave Dynamic Differential Logic

The cornerstone of a secure digital design flow is a logic style with constant power consumption. The power consumption of traditional standard cells and logic is dependent on the signal activity. It depends on both the signal values and the signal transitions, i.e. the Hamming distance between consecutive data values. This is the fundamental reason that information can leak through the power supply and power attacks are possible. To address this problem several logic families have been introduced: asynchronous logic [6], Sense Amplifier Based Logic [8] and Wave Dynamic Differential Logic [9]. They all employ some form of dynamic differential logic, sometimes also referred to as dual rail with precharge logic. In this logic, every signal transition is represented with a switching event, in which the logic gate charges a capacitance, even when the input signals remain constant.

The major disadvantage of the asynchronous approach [6] is that it is extremely difficult. The methodology and tools for the design of large asynchronous logic circuits substantially lags behind that of synchronous circuits. Compared to EDA support for synchronous designs, which is very mature, there is still a shortage of CAD tools to support asynchronous circuit designs.

Sense Amplifier Based Logic (SABL) [8] has been conceived to thwart Differential Power Analysis (DPA). It uses advanced circuit techniques to guarantee that the load capacitance has a constant value. SABL completely controls the portion of the load capacitance that is due to the logic gate. The intrinsic capacitances at the differential in- and output signals are symmetric and additionally it discharges and charges the sum of all the internal node capacitances. A major disadvantage is the non-recurrent engineering costs of a custom designed cell library development. SABL also suffers from a large clock load, as is common to all clocked dynamic logic styles.

Wave Dynamic Differential Logic (WDDL) [9] is implemented with static complementary CMOS logic, which is the default logic style used in standard cell libraries. Static CMOS standard cells are combined to form secure compound standard cells, which have a reduced power signature. Instead of pre-charging with a clock signal as is the case for SABL, in WDDL a pre-discharge wave travels through the circuit. In the precharge phase, the inputs to the WDDL gate are set at 0. This puts the output of the gate at 0 and the precharge wave travels over to the next gate of the combinatorial logic. By way of example, Fig. 2 shows the WDDL AOI32 gate and the original static CMOS gate. Our WDDL library contains 128 cells.

WDDL has many advantages. It can be readily implemented from an existing standard cell library. The design flow is fully supported with accurate EDA library files that come directly from the vendor. WDDL also results in a dynamic and differential logic with only a small load capacitance on the precharge control signal and with the high noise margins of static CMOS. Furthermore, since the gates do not precharge in parallel, it also benefits from a low supply current derivative *di/dt* and peak supply current. As a result, the supply bounce, often a critical problem for signal integrity, is lowered.

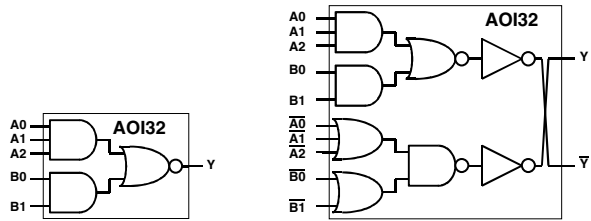

**Fig. 2. AOI32: static CMOS (left); and WDDL (right).**

Building WDDL logic at the logic level will reduce but not eliminate side channel attacks. At the transistor level, the implementation of the WDDL logic also needs to be balanced. It is essential in order to achieve constant power consumption that a fixed amount of charge is used per transition. A standard cell must always charge ideally the same load capacitance. Since a differential logic gate has 2 outputs, of which only 1 switches per event, the load capacitances at the 2 outputs must be matched. The load capacitance consists of the intrinsic in- and output capacitances of the cells and the interconnect capacitances between cells. Additional capacitances can be incorporated inside the compound gates to balance the intrinsic capacitances. Or even custom designed WDDL gates can be made. Yet with shrinking channel-length of the transistors, the share of the interconnect capacitance in the total load capacitance increases and the interconnect capacitances will become the dominant capacitance [10]. Hence, the issue of matching the interconnect capacitances of the signal wires is crucial for the countermeasures to succeed. This is achieved through a special place & route approach.

## 2.2 Multiple Differential Pair Routing

The best strategy to achieve matched interconnect capacitances is to route the output signals with routes that are at all times in parallel and on the same layers. Then independent of the placement, the two routes have the same parasitic effects [11].

The parasitic effects of the interconnects are caused by the resistance of the wires and by the capacitance to the substrate and to wires in upper and lower metal layers. Aside from process variations, these effects are equal for both nets. The resistance is the same since both interconnects have the same length. The capacitances are made the same by the proposed differential routing since in general the length of the differential route is orders of magnitude larger then the distance between the 2 differential routes and one can therefore argue that both nets travel in the same environment.

The pair of interconnects however, need not only to be routed with the same capacitance but also with control over cross-talk. Cross-talk is the phenomenon of noise induced on one wire by a signal switching on a parallel wire. This has an effect on the power consumption. Cross-talk effects are caused by the capacitance to adjacent wires in the same metal layer. Shielding the differential routes on either side with a power or ground line eliminates the cross-talk. Alternatively increasing the distance between the different differential pairs reduces the effect. The tradeoff is an increase in silicon area.

Differential pair routing has been available through gridless routers. But their goal is to route a few critical signals, such as the clock or general reset signal. They are not built for crypto applications where *all* signals need a differential route and thus router performance and completion rate degrade rapidly with increasing number of differential pairs. These tools are unable to route 20K+ differential pairs as an encryption algorithm requires. High-capacity gridded routers on the other hand have no or only limited capability to route differential pairs and often avoid running wires in parallel to prevent cross-talk effects.

We have recently presented a way to work around tool limitations [11]. The technique is built on top of a commercial place & route tool and forces the tool to route the two output signals at all times in adjacent tracks. In the technique, each differential output pair is abstracted as one single "fat" wire. The differential design is routed with the fat wire and at the end the fat wire is decomposed into the differential wire. Fig. 3 demonstrates the place & route approach. The figure shows a placed & routed design consisting of 6 gates. At the left, the result is shown of the fat routing. At the right, the result after decomposition is shown. Two normal wires replace each fat wire.

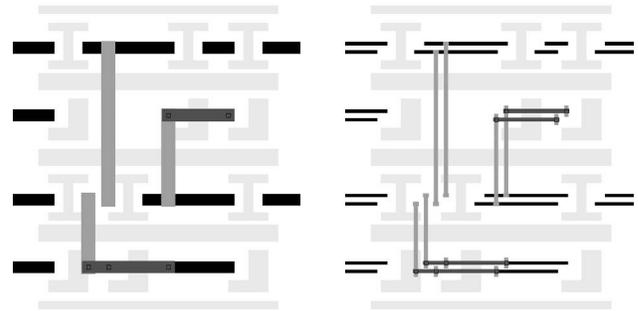

**Fig. 3. Fat design (left); and differential design (right).**

## 2.3 Automated Design Flow

We will now describe the secure digital design flow. Please recall that the design flow is depicted in Fig. 1.

In the logic design phase, the design specifications (`design specs`) are translated into a behavioral model (`behavior.v`). This is the creative phase in the IC design.

Logic synthesis is the process of mapping the behavioral model (`behavior.v`) into logic gates of the library file (`lib.v`). It generates a gate level description of the desired circuit (`rtl.v`). The constraints file (`script`) contains area and timing optimization directives. It also limits the type of gates used during synthesis. The gates available for synthesis depend on the WDDL cells that have been assembled. The library file is the original static CMOS standard cell library file.

The cell substitution procedure modifies the gate level description. A script, e.g. in PERL or in Awk, transforms the gate level netlist (`rtl.v`). Two files are generated: a fat gate level netlist (`fat.v`), which will be used to route the design, and a differential gate level netlist, which will be used in the verification tasks. The differential netlist is obtained by replacing each gate by its WDDL counterpart. The inverters are also removed; the inversions are implemented by switching the nets. The fat netlist is equivalent to the differential netlist except that the differential signals have been abstracted as one single signal. This procedure is not present in a regular design flow. The run time overhead is negligible. The parser required a little less then 4 minutes to generate both files for a prototype IC containing 39000 gates on a SunFire v100 (550MHz., 2GB RAM).

In the place & route step, the fat gate level netlist (`fat.v`) is placed and routed. The place & route tool requires a fat gate library database (`fat_lib.lef`). The library database contains cell macros and routing rules. Information from the original library files is used in procedures such as clock routing and timing driven placement. The resulting design file (`fat.def`) specifies the location of the cells in the core and of the interconnect wires.

The fat to differential routing transformation consists of two separate procedures: (1) a duplication and translation of the fat wires; and (2) a width reduction.

The interconnect decomposition procedure accomplishes the duplication and translation. This procedure edits the fat design file (`fat.def`). In a design file, wire segments are represented by lines between 2 points. The parser duplicates and translates the coordinates of the points. This is the second procedure not present in a regular design flow. It has a negligible timing overhead. The parser required 2 minutes to generate the differential design file (`diff.def`) for our 39K gate IC.

The width reduction is accomplished by updating the library database with the differential library database during the stream out of the design. In this step, the differential design file (`diff.def`) and the differential library database (`diff_lib.lef`), which contains the normal wire definition and the differential gate macros, are combined in the place & route tool to generate the layout (`layout`).

A WDDL gate is composed of gates that are supported by the vendor library. The standard verifications, executed in each stage, accurately verify functionality, delay and area in function of the original gates. Two extra measures validate the inserted cell substitution and interconnect decomposition procedures. A logic equivalence checker, such as Formality [12] or Verplex LEC [13] verifies the equivalence between the fat gate level netlist and the original netlist. Importing the differential gate level netlist during the stream out procedure verifies the equivalence between differential gate level netlist and the differential placed and routed design file.

## 3 Design Example

A test circuit is implemented through the secure digital design flow and through a regular digital design flow using ordinary static CMOS standard cells. The block diagram of the test circuit is depicted in Fig. 4. This circuit has been presented as a sufficient subset of the DES algorithm on which a Differential Power Analysis can be mounted [5]. The algorithm has been reduced to this setup such that the instantaneous supply current transient can be simulated with the transistor level simulator Hspice.

Fig. 5 shows the resulting layouts. The secure implementation and the reference implementation require $12880\mu m^2$ and $3782\mu m^2$ respectively. The WDDL compound gates have been derived from the 0.18μm, 1.8V static CMOS standard cell library that has been used in the regular design flow. The single ended gate level netlist has been obtained through DesignAnalyzer [12]. Place & route have been done in Silicon Ensemble [13] with an aspect ratio of 1 and a fill factor of 80%. The language Awk is used in the parser to generate the fat and differential netlists. The spice netlists, which include the layout parasitics, have been extracted in Virtuoso [13]. In total, 2000 input vectors have been consecutively encrypted with a random input at the plaintext $P_L$ and $P_R$, and with a fixed secret key K, equal to 46. The clock frequency of the circuit is chosen at 125MHz, and 800 'measurement' samples are made per clock cycle.

The clock and input signals are driven by cascaded inverters to provide realistic data and clock signals. The power consumption of the additional input circuitry is excluded from the measurements. The mean energy consumption is 27.1pJ and 4.6pJ for the secure implementa-tion and the reference implementation respectively. The normalized energy deviation, which specifies the absolute range of the variation on the energy consumption per encryption, is 6.6% and 60%. The normalized standard deviation is 0.9% and 12%.

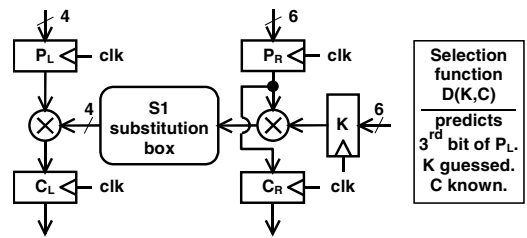

**Fig. 4. Design example: DPA on a module of DES [5].**

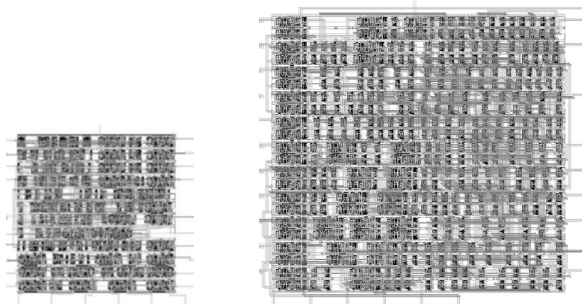

**Fig. 5. Layout with: regular digital design flow (left); and secure digital design flow (right).**

Fig. 6 shows the result from the DPA on the transient SPICE simulation on the extracted netlists, including all parasitic capacitances. Note that in a DPA, the supply current measurements are divided over 2 sets by means of a selection function and a guess on the secret key. The correct secret key is found with the aid of the differential traces. A differential trace is the difference between the typical supply currents of the two sets. The difference will approach zero for a wrong key guess, but has noticeable peaks if the correct secret key has been predicted.

The resistance against DPA is quantified with the MTD, the number of measurements to disclosure. Fig. 6 (left) shows that for the reference design 250 measurements are sufficient to disclose the secret key. The secure digital design flow on the other hand has been effective in reducing the peaks of the differential trace of the correct secret key: the peak-to-peak value of the secret key is conforming with the peak-to-peak value of the other key guesses. The Differential Power Analysis does not reveal the secret key. Fig. 6 (right) shows the peak-to-peak value of the differential traces of the secret key guesses for 2000 measurements. The secret key clearly stands out for the reference implementation.

Perfect security does not exist. Despite the fact that for instance noise and measurement errors will reinforce the level of security, increasing the number of samples may expose the secret key of the secure design. The more powerful an attacker is, the better his results may be. On the other hand, the more one is willing to invest, the higher the security will be. Shielded lines or a larger pitch, balanced intrinsic capacitances or custom designed cells, etc. will improve the security. With the introduction of the secure digital design flow, the problem has been reduced to a problem of parasitic effects.

## 4 Secure Digital Design Flow and other SCAs

A secure digital design flow must create a system that is secure against the collection of known practical side-channel attacks. In the preceding sections, we have dealt with power attacks. We will now discuss how the design flow creates a system that thwarts timing attacks, Electromagnetic Analysis and Differential Fault Analysis.

### 4.1 Timing Attacks

Timing attacks are a class of cryptanalysis that takes advantage of timing information [1]. For instance, the arrival time of the ciphertext may be determined by operations that depend on the secret key. Making the power consumption uniform does not remove the threat of a timing attack. Conditional branching with unequal lengths for example can still be seen. Therefore, a careful design that at all times has a worst case running time is still necessary.

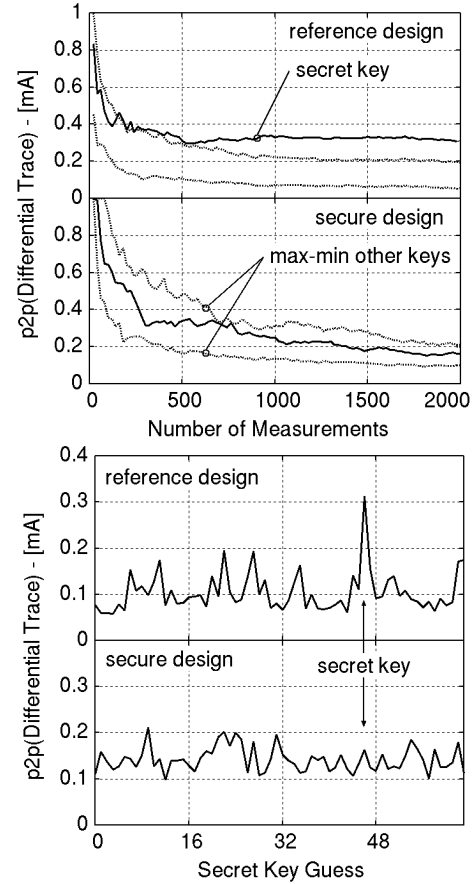

Fig. 6. Measurements to disclosure (top); peak-to-peak of differential traces at 2000 measurements (bottom).

Power measurements can nevertheless provide substantial timing information in spite of this conventional design practice. They expose idle cycles, which have been inserted to hide conditional branches with unequal lengths. If the state of the circuit does not change, the gates of a regular design will not switch and do not dissipate any power. This is not the case for a system that employs a logic style with a 100% switching factor, as is done in the secure digital design flow. When desired, idle cycles can be inserted. Every gate has a switching event in every cycle, whether or not useful data is processed or in other words whether or not idle cycles have been inserted.

### 4.2 Electromagnetic Analysis

The flow of electric charges produces an electromagnetic field. Electromagnetic Analysis (EMA) is the equivalent of a power attack but instead uses the electromagnetic fields generated by the (dis)charging gates as the side channel information [3]. There are differences. A power attack has only access to the global power consumption. An EMA however, can do measurements that are confined to a small area of the security IC. On the other hand, a power attack can be mounted rather quickly with off-the-shelf devices, whereas an EMA requires an advanced measurement setup.

The electromagnetic fields are generated by the electric charge that flows through the power lines and either of the 2 differential output wires. Since ideally the same amount of charge is used, the only option to differentiate between different switching events is to detect which of the 2 differential output wires the charge flows through.

The differential output wires have been routed at a distance of about 1μm apart with the special place & route approach, while their length is 10 to several 100μm. The measurement probe on the other hand is positioned at a few millimeters from the device [3]. The resulting measurement setup is depicted in Fig. 7. The measurement probe must distinguish between electromagnetic fields generated between two antennas for which the distance in between (1μm) is orders of magnitude smaller then their length (10-100μm), while the distance between the probe and the antennas (1-10mm) is another multiple order of magnitude larger. Furthermore, many cells switch simultaneously and thus large numbers of antennas are broadcasting at the same time and in the same area. In the available literature [3],[14],[15], a measurement setup with this kind of accuracy has not been demonstrated.

### 4.3 Differential Fault Analysis

In Differential Fault Analysis (DFA), the attacker tries to force an error in the internal state of the circuit, and subsequently exploit weaknesses of the algorithm under malfunctioning. The main countermeasure against DFA is fault detection. The idea is to monitor if one has tried to corrupt the computation, and in such an event to shut down the processor and delete any valuable information as soon as possible.

A dynamic and differential logic style, such as WDDL, employs redundant encoding. This means that a state is actually represented by 2 bits instead of 1 bit, as is the case in a normal single ended logic, such as static CMOS logic. With redundant encoding error detection is possible. For instance, in a glitch attack, the clock frequency is temporarily increased with the intention to force a state-bit or a selected conditional branch by not giving enough time for all calculations to complete. If at the rising clock edge an input signal to a register is differential it means that it has correctly evaluated. On the other hand if both inputs remain at zero, the clock frequency has been increased and the circuit should be put in the alarm state. Of course, it is sufficient to monitor the critical path of the system. In other DFAs, electromagnetic radiation, such as light or radio waves, is used to flip a state-bit. Here a similar approach can be worked out, except that the detection is not restricted to 1 register.

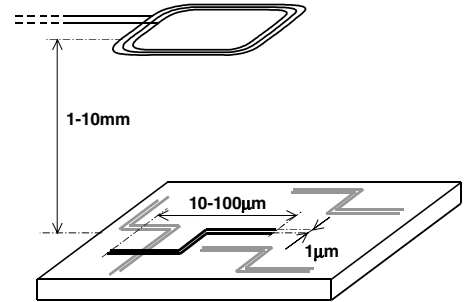

**Fig. 7. Measurement setup (not drawn to scale).**

## 5 Conclusions

We have presented a secure digital design flow. The design flow provides an accessible means to fabricate a security IC that is SCA resistant regardless of the implementation details. It relies on a logic style that has constant power consumption and a place & route approach that controls the parasitic effects. The design flow is completely supported by mainstream EDA tools and depending on the required level of security several options are available to strengthen the security. Experimental results have demonstrated that the design flow results in a DPA resistant layout and we have argued the effective obstruction of a collection of know practical SCAs.